# First-principles study of photovoltaic and thermoelectric properties of AgBiSCl$_2$[*]

WANG Sihang[1], CHEN Menghan[2], ZHANG Liping[1],

1.School of Materials Science and Physics, China University of Mining and Technology, Xuzhou 221116, China

2.School of Mechanics and Civil Engineering, China University of Mining and Technology, Xuzhou 221116, China

**Abstract**

This work systematically investigates the potential of the hybrid anion semiconductor AgBiSCl$_2$ for photovoltaic and thermoelectric applications, aiming to provide theoretical guidance for high-performance energy conversion devices. Structural analysis reveals favorable ductility and a relatively low Debye temperature (219 K). Electronic structure calculations show that AgBiSCl$_2$ is a direct band gap semiconductor, with a gap of approximately 1.72 eV after including spin-orbit coupling effects. The conduction band is mainly derived from Bi 6p orbitals, while the valence band is dominated by contributions from Ag 4d, Cl 3p, and S 3p orbitals.

Analysis of interatomic interactions indicates that Ag-S and Ag-Cl bonds are relatively weak, resulting in local structural softness and enhanced lattice anharmonicity. These weak bonds facilitate phonon scattering and give rise to low-frequency localized "rattling" vibrations primarily associated with Ag atoms, contributing to reduced lattice thermal conductivity. In contrast, Bi—S bonds exhibit stronger, directional interactions, which help stabilize the overall structure. The coexistence of weak bonding and strong lattice coupling enables favorable modulation of thermal transport properties.

Optically, AgBiSCl$_2$ possesses a high static dielectric constant ($\varepsilon_1(0)$ = 5.60) and exhibits strong absorption in the ultraviolet region, with absorption coefficients rapidly exceeding $1 \times 10^6$ cm$^{-1}$. A theoretical solar conversion efficiency of up to 28.06% is predicted for a 3 μm-thick absorber layer, highlighting its potential as a high-performance photovoltaic material.

In terms of thermal transport, phonon spectrum exhibit mode hardening with temperature increasing, while flat optical branches in the 30-70 cm$^{-1}$ range enhance phonon scattering.

The localized Ag vibrations intensify the anharmonicity, reducing phonon lifetimes and group velocities. As a result, at 300 K, the lattice thermal conductivities via the Peierls and coherent channels are calculated to be 0.246 W·m$^{-1}$·K$^{-1}$ and 0.132 W·m$^{-1}$·K$^{-1}$, respectively. For electronic transport, the p-type material maintains a higher Seebeck coefficient than the n-type, while the latter shows greater electrical conductivity. At 700 K, the thermoelectric figure of merit (*ZT* ) reaches 0.77 for p-type and 0.69 for n-type AgBiSCl$_2$, indicating promising high-temperature thermoelectric performance.

In summary, AgBiSCl$_2$ exhibits excellent potential for dual photovoltaic and thermoelectric applications. Its unique bonding features and lattice response mechanisms offer valuable insights into designing multifunctional energy conversion materials.

## 1. Introduction

Mixed anion compounds are new material systems in which two or more anions (such as oxyhalides [1], oxysulfides [2], and sulfur halides [3]) are simultaneously introduced into the same crystal framework. Compared with traditional single anion compounds, the diversity of anion types and proportions endows them with extremely rich potential for structure and property control. By designing properties such as charge, ionic radius, electronegativity, and polarizability of different anions, comprehensive regulation of bandgap width, electron transport, optical absorption, and magnetic behavior can be achieved [4]. In addition to their attractive physical properties for optical applications, mixed-anion compounds are also promising as low-thermal-conductivity materials [5]. Such materials exhibit low thermal conductivity and can play an important role in the field of thermoelectric. The efficiency of thermoelectric materials is expressed by the dimensionless figure of merit (thermoelectric figure of merit, *ZT*):

$$ZT = \frac{S^2 \sigma T}{\kappa}, \tag{1}$$

Where $S$, $\sigma$ and $T$ are Seebeck coefficient, electrical conductivity and absolute temperature, respectively; Thermal conductivity $\kappa = \kappa_L + \kappa_e$, $\kappa_L$ is the lattice thermal conductivity, $\kappa_e$ is the electronic thermal conductivity. To obtain a high *ZT* value, thermoelectric materials need to have a high power factor (PF = $S^2\sigma$) and a low thermal conductivity. Cu-based or Ag-based compounds, as a typical representative, use metal-anion weak coordination bonding to induce strong anharmonic phonon dynamics and atomic "rattling" vibration[6], which suppresses the propagation of transverse phonons to a certain extent, resulting in extremely low lattice thermal conductivity, such as $Cu_3SbSe_3$[7], $AgCrSe_2$[8] and other systems. In recent years, Ruck et al.[9] synthesized black plate-like crystals of $AgBiSCl_2$ by thermal reaction of equimolar amounts of AgCl and BiSCl. Quarta et al.[10] obtained mixed anion compound $AgBiSCl_2$ nanocrystals by colloidal synthesis method, and studied its crystal structure, optical properties and electronic structure. It is found that the material has direct band gap and photo luminescence properties, which provides a new direction for its application in the field of optoelectronics. However, there is a lack of detailed research on the correlation of crystal structure, chemical bonding, optical absorption, electrical transport and thermal transport properties of these compounds, which to some extent limits their application in related optoelectronic and thermoelectric fields. In this paper, the potential application value of $AgBiSCl_2$ material is deeply explored. The performance of $AgBiSCl_2$ in thermoelectric and photoelectric properties is studied by first principles combined with Boltzmann transport theory, and the laws of electrical transport and thermal transport are deeply analyzed, which provides a solid theoretical basis and guidance for the application of $AgBiSCl_2$ in photoelectric detection, solar cells and thermoelectric power generation devices.

## 2. Details of calculation

Density functional theory (DFT) was employed, implemented using the Vienna ab initio simulation package (VASP) code [11]. The exchange-correlation potential was treated within the Perdew-Burke-Ernzerhof (PBE) generalized gradient approximation (GGA) form [12], and the wave functions were described using the projector-augmented wave (PAW) method [13]. During structural optimization, a cutoff energy of 500 eV was chosen for the plane-wave basis set. A Gamma-centered 12×4×6 k-point grid was used. The convergence criteria for energy and forces were set to $1\times10^{-8}$ eV and $1\times10^{-6}$ eV/A°, respectively. The hybrid functional (Heyd-Scuseria-Ernzerhof, HSE06) was adopted to obtain reliable bandgap and optical properties [14]. Furthermore, spin-orbit coupling (SOC) effects were considered for the calculations of band structure, optical, and electrical transport properties due to the presence of heavy elements.

The optical properties of the material are described by the real part $\varepsilon_1(\omega)$ and the imaginary part $\varepsilon_2(\omega)$ of the complex dielectric function, where $\varepsilon(\omega) = \varepsilon_1(\omega) + i\varepsilon_2(\omega)$. The imaginary part $\varepsilon_2(\omega)$ is related to the energy absorption of the material and can be calculated using the following formula [15]:

$$\varepsilon_2(\omega) = \frac{Ve^2}{2\pi\hbar m_0^2 \omega^2} \int d^3k \sum_{nn'} |\langle kn|p|kn'\rangle|^2 f(kn)$$
$$\times (1 - f(kn'))\delta(E_{kn} - E_{kn'} - \hbar\omega), \quad (2)$$

In the formula, $e$ represents the charge, $m_0$ is the electron mass, $V$ denotes the unit cell volume, $\hbar$ is the reduced Planck constant, $\langle kn|p|kn'\rangle$ represents the momentum transition matrix element, $kn$ and $kn'$ are the wave functions of the valence band and conduction band, respectively, $f(kn)$ and $f(kn')$ represent the Fermi–Dirac occupation numbers of wave vector ***k*** in bands $n$ and $n'$, $E_{kn}$ and $E_{kn'}$ are the corresponding energy eigenvalues, and $\hbar\omega$ is the energy of the incident photon. The real part $\varepsilon_1(\omega)$ represents the material's ability to store electric field energy, indicating the energy consumption during the formation of an electric dipole. It is closely related to interband transitions and can reflect the extent of electron transitions in the material, which can be calculated by the following expression [16]:

$$\varepsilon_1(\omega) = 1 + \frac{2}{\pi}M \int_0^\infty \frac{\omega' \varepsilon_2(\omega')}{\omega'^2 - \omega^2} d\omega', \quad (3)$$

where $M$ denotes the principal value of the integral. Once the dielectric function is obtained, optical parameters such as the optical absorption coefficient, reflectivity, and refractive index can be calculated.

The electrical transport properties are calculated using the AMSET software package [17]. This method employs the relaxation time approximation and accounts for scattering mechanisms including acoustic deformation potential (ADP) scattering, ionized impurity (IMP) scattering, and polar optical phonon (POP) scattering. The relaxation time is determined by Matthiessen's rule [18]:

$$\frac{1}{\tau} = \frac{1}{\tau^{ADP}} + \frac{1}{\tau^{IMP}} + \frac{1}{\tau^{POP}}, \quad (4)$$

Where $\tau^{ADP}$, $\tau^{IMP}$, and $\tau^{POP}$ are the relaxation times of ADP, IMP, and POP scattering, respectively. The differential dispersion rate is calculated using Fermi's golden rule [19]:

$$\tau^{-1}_{nk \to mk+q} = \frac{2\pi}{\hbar}|g_{nm}(\mathbf{k},\mathbf{q})|^2 \delta(\varepsilon_{nk} - \varepsilon_{mk+q}), \quad (5)$$

Where $n$ and $m$ represent the band numbers of the initial and final States, respectively; $k$ represents the wave vector of the initial electron, and $q$ represents the momentum transfer in the scattering process; The $n\boldsymbol{k}$ indicates that the electron is in the initial state with band $n$ and wave vector $\boldsymbol{k}$, while the $m\boldsymbol{k}+\boldsymbol{q}$ indicates that the electron is in the final state with band $m$ and wave vector $\boldsymbol{k}+\boldsymbol{q}$ after scattering; $\varepsilon$ and $g$ are the carrier energy and coupling matrix element, respectively; $\delta$ is the Dirac delta function, which ensures energy conservation during the scattering process. The potential basis elements for the ADP, IMP and POP scattering processes are given[20] by

$$g_{nm}^{\mathrm{ADP}}(\boldsymbol{k},\boldsymbol{q})=\left(\frac{k_{\mathrm{B}}T\alpha_v^2}{B_0}\right)^{1/2}\left\langle\psi_{mk+q}\middle|\psi_{nk}\right\rangle, \qquad (6)$$

$$g_{nm}^{\mathrm{IMP}}(\boldsymbol{k},\boldsymbol{q})=\left(\frac{n_{ii}Z^2e^2}{\varepsilon_0}\right)^{1/2}\frac{\left\langle\psi_{mk+q}\middle|\psi_{nk}\right\rangle}{|\boldsymbol{q}|^2+\beta^2}, \qquad (7)$$

$$g_{nm}^{\mathrm{POP}}(\boldsymbol{k},\boldsymbol{q})=\left[\frac{\hbar\omega_{\mathrm{po}}}{2}\left(\frac{1}{\varepsilon_\infty}-\frac{1}{\varepsilon_0}\right)^{1/2}\right]\frac{\left\langle\psi_{mk+q}\middle|\psi_{nk}\right\rangle}{|\boldsymbol{q}|}, \qquad (8)$$

Where $k_\mathrm{B}$, $\alpha_v$, $B_0$, $n_{ii}$, $\beta$, $Z$, $\varepsilon_\infty$, $\varepsilon_0$ and $\omega_{\mathrm{po}}$ are Boltzmann constant, deformation potential, bulk modulus, ionized impurity concentration, inverse of screening radius, number of doped charges, high frequency dielectric constant, static dielectric constant and polarization optical branch frequency, respectively.

For phonon thermal transport properties, the interatomic force constants (IFCs) [21,22] were calculated by ALAMODE software package, and 3 × 1 × 1 supercell was used to calculate IFCs. The second-order force constants were extracted by the finite displacement method, that is, in each configuration, a single atom was displaced by 0.01 Å from the equilibrium position in a certain direction, and then the second-order force constants were obtained by least squares fitting. High-order IFCs are generated by compressive sensing lattice dynamics (CSLD)[23]. *ab initio* molecular dynamics (AIMD) simulations were performed at 300 K for 2000 steps, and random displacements of 0.1 Å were applied to all atoms in the trajectory at equal intervals for 10 structures to obtain multiple random structures, and force-displacement data sets were obtained by single-point energy calculation for these configurations. The Taylor expansion potential including up to the sixth order anharmonic terms is fitted using an elastic net. The higher-order force constants take into account the interaction between all atoms, and the truncation radius of 12 (in Bohr units) is set from the third order to the sixth order, with a fitting relative error of 5%. The convergence of the lattice thermal conductivity is shown in Figure S1 (https://wulixb.iphy.ac.cn/article/doi/10.7498/aps.74.20250650), and the lattice thermal conductivity is finally calculated using an 11 × 11 × 11 $q$ point density grid.

For a specific phonon mode, its contribution to the thermal conductivity can be divided into two parts: particle-like transport and wave-like transport. Recently, it has been found that the coherent transport of thermal conductivity (wavy diffusion) is not only the main thermal conduction in amorphous crystals, but also the excitation in complex crystals, such as $AgTlI_2$[24], $Cs_2AgBiBr_6$[25], $BaAg_2SnSe_4$[26], $A_2PdPS_4I$ ($A$ = K, Rb, Cs)[27], etc. The two-channel lattice thermal conductivity is the sum of the two parts of thermal conductivity, i.e., $\kappa_L = \kappa_p + \kappa_c$. Where $\kappa_p$ takes into account the contribution of the diagonal terms ($j=j'$) of the phonon group velocity operator to the heat transport. $\kappa_p$ can be written as[28]:

$$\kappa_p^{\mu\nu} = \frac{1}{VN_q} \sum_{q,j} c_{qj}(T) v_{qj}^\mu v_{qj}^\nu \tau_{qj}(T), \qquad (9)$$

Where the superscripts $\mu$ and $\nu$ denote spatial components (such as $x$, $y$ or $z$), $V$ is the elementary cell volume, $N_q$ is the number of wave vectors, $q$ is the phonon wave vector, $j$ is the number of phonon branches, $c_{qj}$ represents the contribution of phonon modes ($q$, $j$) to the specific heat, $v$ is the group velocity, and $\tau$ is the phonon lifetime. The $\kappa_c$ and are called the thermal conductivity of coherence, which reflects the wave-like thermal transport characteristics and originates from the interference between the vibrational eigenstates $j$ and $j'$, which is called the thermal conductivity of coherence. Based on the Wigner transport equation, the $\kappa_c$ expression is [29]

$$\kappa_c^{\mu\nu}(T) = \frac{1}{VN_q} \sum_{\substack{q,jj' \\ j \neq j'}} \frac{c_{qj}\omega_{qj'} + c_{qj'}\omega_{qj}}{\omega_{qj} + \omega_{qj'}} \\ \times v_{qjj'}^\mu v_{qjj'}^\nu \frac{\Gamma_{qj} + \Gamma_{qj'}}{(\omega_{qj} - \omega_{qj'})^2 + (\Gamma_{qj} + \Gamma_{qj'})^2}. \qquad (10)$$

Where $c_{qj'}$ represents the contribution of the phonon mode ($q$, $j'$) to the specific heat; $\omega_{qj}$ and $\omega_{qj'}$ denote the corresponding phonon frequencies; The $\Gamma_{qj}$ and $\Gamma_{qj'}$ represent the corresponding phonon linewidth, reflecting the phonon lifetime and scattering intensity. For a specific phonon, its contribution to the thermal conductivity can be divided into two parts: phonon transport and coherent transport, and the proportion of the two parts is determined by the contrast between the phonon lifetime and the Wigner limit. The Wigner limit is $3N_{at}/\omega_{max}$, where $N_{at}$ is the number of atoms in the primitive cell and $\omega_{max}$ is the maximum phonon frequency; The Ioffe-Regel limit $\tau_{Ioffe-Regel}$ is $1/\omega$, where $\omega$ is the phonon frequency. The contribution of different frequency phonons to $\kappa_p$ and $\kappa_c$ is analyzed by Wigner limit and Ioffe-Regel limit. When the phonon lifetime $\tau > \tau_{Wigner}$ is particle-like, the phonon determines the size of $\kappa_p$, and the phonon lifetime at $\tau_{Ioffe-Regel} < \tau < \tau_{Wigner}$ contributes to $\kappa_c$, which is wave-like. When the phonon lifetime $\tau$ is close to $\tau_{Wigner}$, the two mechanisms contribute comparably[29,30].

## 3. Results and Discussion

### 3.1 Electronic structure

As shown in the Fig.1(a), the AgBiSCl$_2$ crystal contains 4 Bi atoms, 4 Ag atoms, 8 Cl atoms, 4 S atoms. The lattice parameters of the unit cell, $a$ = 4.07 Å, $b$ = 14.17 Å and $c$ = 8.69 Å, are close to the experimental data (see Supplementary Material Table S1 (https://wulixb.iphy.ac.cn/article/doi/10.7498/aps.74.20250650)) and belong to the orthorhombic *Cmcm* space group. Ag$^{1+}$ is coordinated by two equivalent S$^{2-}$ atoms and four equivalent Cl$^{1-}$ atoms, forming an angle-twisted and edge-shared AgS$_2$Cl$_4$ octahedral structure. The angle-sharing octahedron has a tilt angle of 58°. Both Ag - S bond lengths are 2.46 Å, and all Ag - Cl bond lengths are 3.09 Å. Bi$^{3+}$ is coordinated by two equivalent S$^{2-}$ atoms and six equivalent Cl$^{1-}$ atoms in an 8-coordination geometry. The two Bi - S bond lengths are both 2.71 Å, and two of the Bi - Cl bond lengths are shorter (2.81 Å) and four are longer (3.21 Å). S$^{2-}$ is coordinated by two equivalent Ag$^{1+}$ atoms and two equivalent Bi$^{3+}$ atoms, forming a distorted angle-shared SAg$_2$Bi$_2$ tetrahedron. Cl$^{1-}$ is coordinated by two equivalent Ag$^{1+}$ atoms and three equivalent Bi$^{3+}$ atoms in a 5 - coordination geometry. The calculated mechanical properties of the material are listed in Table S1 (https://wulixb.iphy.ac.cn/article/doi/10.7498/aps.74.20250650). The ratio of the bulk modulus to the shear modulus of the material is 2.19, and the Poisson's ratio is 0.30, indicating that the material has good ductility and flexibility. In addition, AgBiSCl$_2$ has a very low Debye temperature of 219 K. According to the Slack formula [31], the low Debye temperature implies that the material may have a low lattice thermal conductivity.

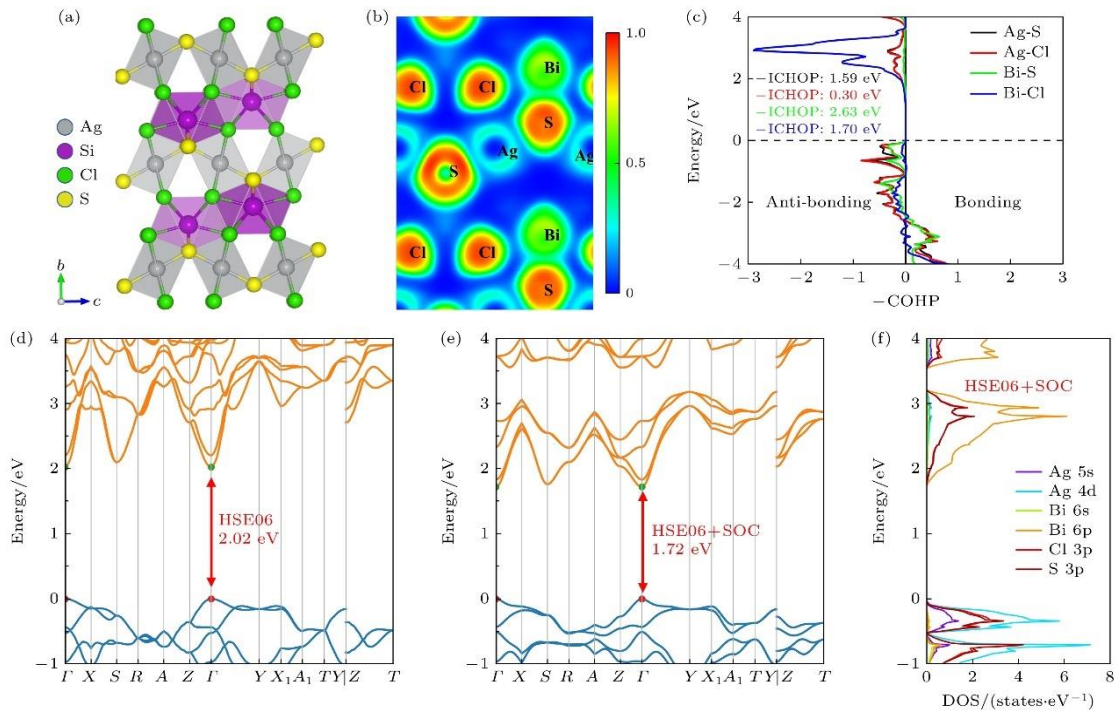

**Figure 1.** (a) Crystal structure of AgBiSCl$_2$; (b) electron localization function; (c) COHP analysis of nearest-neighbor atom pairs; (d) band structure (HSE06); (e) band structure (HSE06 + SOC); (f) density of states.

Figure 1(b) shows the electron localization function (ELF) projected along the *a* direction. The closer the filled color is to blue, the closer the ELF is to 0, indicating that the electrons tend to be delocalized in this region. The closer the filled color is to red, the closer the ELF is to 1, indicating that the electrons tend to be localized in this region. The local electron density around Ag is very low, which is significantly lower than that of other atoms, indicating that S and Cl elements show much stronger electron attraction than Ag, indicating that Ag atoms are more delocalized and do not form bonds with octahedral. It has been found that Cu atoms in CuBiSeCl$_2$ also have similar delocalization phenomena [32], and there is a polar covalent bond between Bi and S atoms (ELF is 0.5). The crystal orbital Hamilton populations (COHP) analysis was performed using LOBSTER software. The calculated negative COHPs of different nearest-neighbor atom pairs are shown in the Fig. 1(c). The negative COHP of the bonding state is positive, and the negative COHP of the antibonding state is negative. Ag — S/Cl has an obvious antibonding state near the valence band below the Fermi level. The antibonding state below the Fermi level weakens the chemical bond and increases the lattice anharmonicity. The negative ICHOP (integration of COHP in the energy range below the Fermi level) values in the figure show that the negative ICOHP of Ag-S/Cl is less than that of Bi-S/Cl, indicating that the bonding of Bi-S/Cl is stronger, while the bonding of Ag- S/Cl is weaker.

Thermoelectric properties strongly depend on the accuracy of the band gap and the electronic States near the band edge. Fig. 1(d),(e) is the band structure of AgBiSCl$_2$. In the figure, both the valence band maximum (VBM) and the conduction band minimum (CBM) are located at the $\varGamma$ point, indicating that the material is a direct gap semiconductor. Using the HSE06 hybrid functional, the band gap without spin-orbit coupling is 2.02 eV, and the band gap with spin-orbit coupling is 1.72 eV. The Fig. 1(f) is the density of States. It can be seen that the bottom of the conduction band is mainly composed of Bi 6p and S 3p orbitals, and the top of the valence band is mainly contributed by Ag 4D, Cl 3p and S 3p orbitals. The valence band is flatter than the conduction band, indicating that holes have a larger effective mass, as listed in the Tab. 1. According to reference [33], the Seebeck coefficient of p-type AgBiSCl$_2$ is larger than that of n-type, but the conductivity is smaller.

Table 1. Effective masses of holes and electrons of AgBiSCl$_2$ ($m_0$ = 9.1 × 10$^{-31}$ kg).

| Carrier type | $m_x/m_0$ | $m_y/m_0$ | $m_z/m_0$ |
|---|---|---|---|
| h | 0.873 | 0.873 | 0.666 |
| e | 0.274 | 0.274 | 0.537 |

**3.2 Optical property**

The optical properties of AgBiSCl$_2$ are shown in the Fig. 2, and the Fig. 2(a) is the real part of the dielectric function, which is related to the energy polarization. This analysis is essential for understanding the interaction of materials with light in solar cell devices. When the frequency is zero, the real part of the dielectric function is called the static dielectric constant$\varepsilon_1$ (0), and the values in each direction are listed in Tab. 2. With the increase of photon energy, the maximum values are 9.83, 7.26 and 8.72 in the directions of *a*,*b* and *c*, respectively. With the increase of photon energy, these values gradually decrease and become negative. The imaginary part of the dielectric function reveals that the material absorbs light. As shown in Fig. 2(b), there is almost no absorption at energies less than 1.72 eV, indicating that the optical band gap value is equal to the fundamental band gap value, which is related to the direct interband transition of electrons between the top of the valence band and the bottom of the conduction band. The imaginary part of the dielectric function has several extreme values in the range of 3- 8 eV, and the larger the value, the better the absorption properties. The first extreme is called the first absorption peak, which is the result of the electronic transition between the occupied state and the unoccupied state near the Fermi level. The first absorption peaks in the *a*, *b* and *c* directions are taken at 3.45, 5.09 and 4.78 V, respectively. Fig. 2(c) is the optical absorption coefficient, and the change is as expected, there is no optical absorption before 1.72 eV, which is of the order of 1 × 10$^5$ cm$^{-1}$ in the visible range; Then it reaches a maximum in the ultraviolet region, and its absorption coefficient is higher than 1 × 10$^6$ cm$^{-1}$ for each direction. The high absorption peak in this region indicates that AgBiSCl$_2$ is suitable for ultraviolet detectors. Reflectivity *R* and refractive index *n* are two important parameters required for solar energy applications. Reflectivity gives a measure of reflected light or radiation. The refractive index reflects the transparency of the material. Reflectivity is shown as Fig. 2(d). It can be clearly seen that *R* and *n* have a definite value at frequency zero, called the static values *R* (0) and *n* (0), which are also listed in Tab. 2. The *R* is low in the visible light range, and combined with the low absorption coefficient near the red light band, the *R* can be used as a window or protective layer material of a red light laser, the low absorption reduces the thermal effect, and the low reflection reduces the power loss. Fig. 2(e) is the refractive index, which maintains a high value greater than 2 in both the visible and near-UV ranges, with a maximum value of 3.22, and is achieved at an energy of 3.06 eV in the *a* direction. The high refractive index makes it possible to be used in optical waveguide devices, which is helpful to achieve good optical waveguide effect.

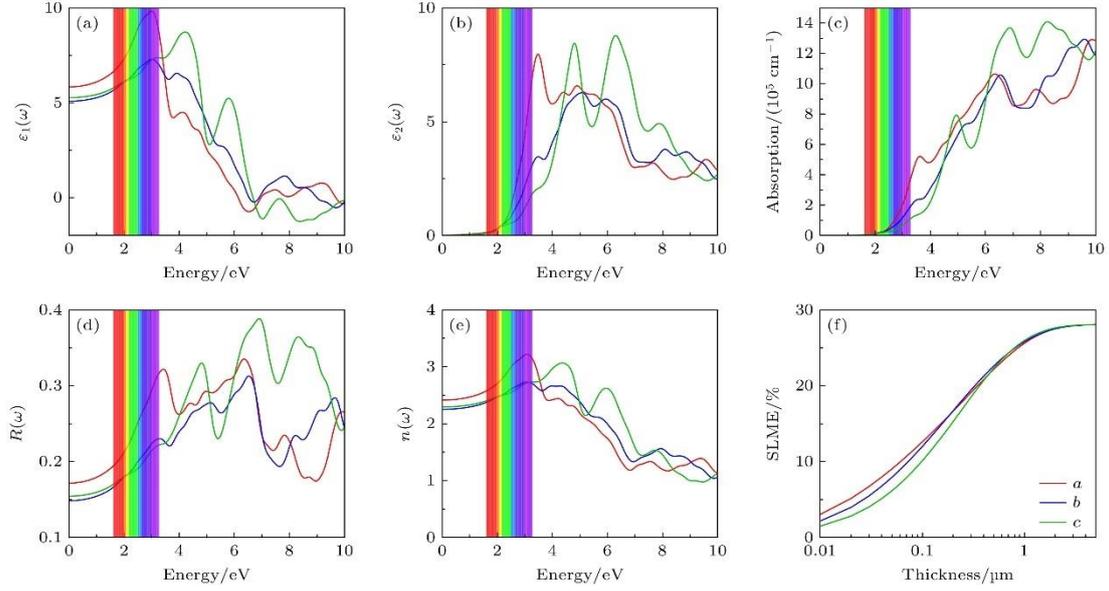

**Figure 2.** (a) Real part of the dielectric function; (b) imaginary part of the dielectric function; (c) optical absorption coefficient; (d) reflectivity; (e) refractive index; (f) SLME.

Table 2. Calculated zero-frequency dielectric constant, reflectivity, and refractive index of AgBiSCl$_2$.

| Parameters | a | b | c | Average |
|---|---|---|---|---|
| $\varepsilon_1(0)$ | 5.84 | 5.08 | 5.27 | 5.40 |
| $R(0)$ | 0.17 | 0.15 | 0.15 | 0.16 |
| $n(0)$ | 2.42 | 2.25 | 2.3 | 2.32 |

The conversion efficiency of photovoltaic devices is one of the key indicators to evaluate their performance, which is restricted by many factors, such as thermodynamic properties, material defects, optical properties and preparation processes. In order to compare the photoelectric conversion potential of different materials, the spectroscopic limited maximum efficiency (SLME) model was introduced. The model was proposed by Yu *et al.* [34], which estimates the solar absorption capacity of materials through two basic parameters of band gap and absorption coefficient, and then predicts their theoretical efficiency. In practical applications, SLME is often used to screen and evaluate the effectiveness of absorber materials. For example, Zhu et al. Calculated the SLME values of various lead-free double perovskite materials [35], and predicted four kinds of perovskite batteries with SLME exceeding 23%. The Fig. 2(f) show the trend of SLME as a function of absorber thickness, measured in μm. When theof AgBiSCl$_2$ is 3 μm, the maximum SLME of the three directions is 28. 06%, which is higher than that of many famous chalcogenide perovskites, such as *β* - CaZrSe$_3$ (23. 08%), SrZrSe$_3$ (22. 11%), CaHfSe$_3$ (20. 09%) and SrHfSe$_3$ (20. 61%) [36]. To sum up, the high refractive index, high light absorption coefficient and excellent SLME characteristics together indicate that $AgBiSCl_2$ is a potential photoelectric absorption layer material, which can provide a reference for the design and development of the next generation of high-efficiency photovoltaic devices.

**3.3 Anharmonic lattice dynamics.**

Strong anharmonicity leads to phonon frequency shift and phonon broadening [37,38], which results in very low thermal conductivity of materials. The Fig. 3(a) is the effect of higher order anharmonicity on phonon dispersion at different temperatures. In this paper, the self-consistent phonon (SCPH) theory is used to deal with the thermal transport process of $AgBiSCl_2$. When the temperature increases from 100 K to 700 K, the phonon hardening occurs in $AgBiSCl_2$. In addition, there are many flat optical branches in the phonon frequency range from 30 to 70 $cm^{-1}$, which can produce a large number of scattering channels to effectively scatter phonons [39]. The Fig. 3(b) are the dispersion curves of the three acoustic branches and the three low-frequency optical branches at zero temperature. The results show that there are multiple representative avoided crossings between $TO_1$ and LA, as indicated by the black circles. Figure S2 (https://wulixb.iphy.ac.cn/article/doi/10.7498/aps.74.20250650) shows the vibration visualization of the LA branch at the *R* point, and it can be seen that the amplitude of the Ag atom is much larger than that of the other atoms. The avoided crossing point is believed to be associated with phonon rattling vibrations and low lattice thermal conductivity [40]. The rattling vibration is a specific physical feature of "phonon liquid, electron crystal", which can enhance phonon scattering and reduce lattice thermal conductivity [41]. As shown by the projected phonon density of States in Fig. 3(c), the low-frequency phonon modes below 30 $cm^{-1}$ are mainly contributed by Ag atoms, the high-frequency phonon modes around 200 $cm^{-1}$ are mainly contributed by Cl atoms, and the phonon modes above 200 $cm^{-1}$ are dominated by S atoms. Fig. 3(d) is the mean square displacement (MSD) of $AgBiSCl_2$ as a function of temperature, which is a measure of the deviation of atomic positions from a reference position over time. In the figure, the displacement of Ag atom is obviously larger than that of other atoms Bi, S and Cl, and is more obvious in the *x* and *y* directions, indicating that the Ag atom is seriously delocalized and soft [42]. In order to further understand the atomic vibration in the crystal, the potential energy curve of each atom of $AgBiSCl_2$ was calculated, as shown in Figure S3 (https://wulixb.iphy.ac.cn/article/doi/10.7498/aps.74.20250650). It can be seen that the Bi atom is confined in a steep potential well, which indicates a relatively strong bond character. Ag atoms are located in a very flat potential well, especially along the *x* direction. When subjected to small perturbations or thermal excitations, Ag atoms rattle from their equilibrium positions, which scatters phonons and reduces the thermal conductivity. To sum up, it can be considered that in $AgBiSCl_2$, the rattling vibration of Ag atoms increases the anharmonicity of the lattice and leads to a low $\kappa_L$.

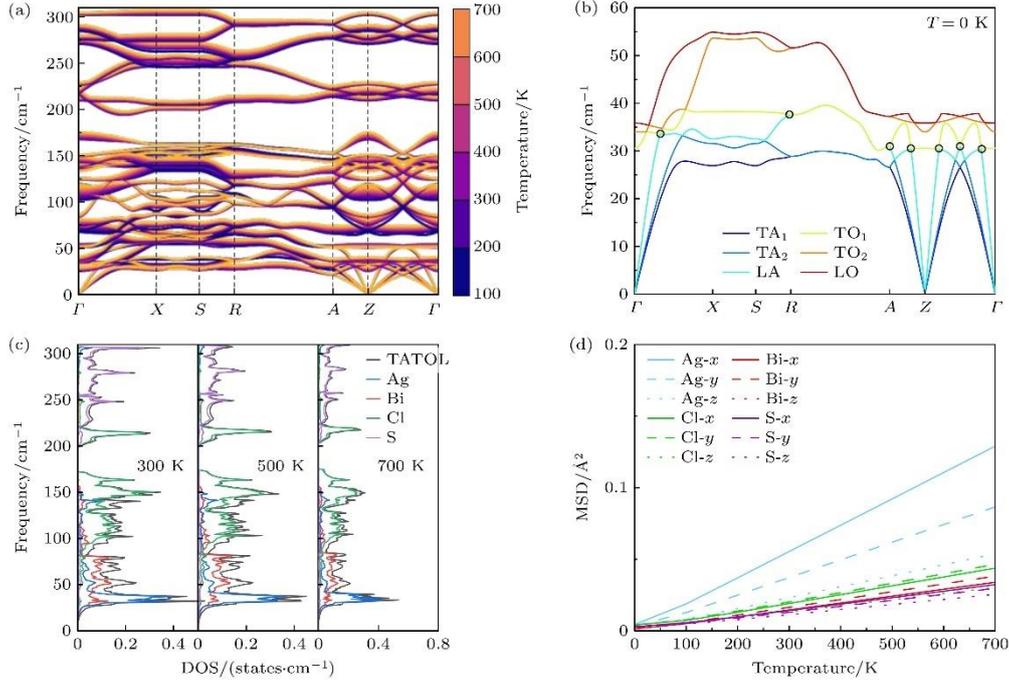

**Figure 3.** (a) Phonon dispersion of AgBiSCl$_2$ at different temperatures; (b) acoustic branches and low-frequency optical branches at 0 K; (c) density of states at 300 K, 500 K, and 700 K; (d) root-mean-square displacement at different temperatures.

Fig. 4(a) shows the $\kappa_p$ and $\kappa_c$ calculated by the two-channel thermal transport model, respectively, with the average $\kappa_p$ and $\kappa_c$ of 0.246 W/ (m · K) and 0.132 W/ (m · K) at 300 K. From the cumulative $\kappa_p$ shown in Fig. 4(b), it can be seen that for the *a* and *c* directions, both low and high frequency phonons contribute significantly to the $\kappa_p$ in AgBiSCl$_2$. However, the $\kappa_p$ in the *b* direction is mainly contributed by the low-frequency phonons in the 0 - 170 cm$^{-1}$, and the thermal conductivity in all directions is very low. Due to the avoided crossing, the phonon dispersion in the low frequency region becomes very flat, resulting in a sharp reduction in the phonon group velocity and a significant weakening of the contribution to the lattice thermal conductivity. At the same time, the avoided crossing causes the original acoustic vibrational mode to mix with the optical mode, increasing the scattering channel and making the lifetime of these low-frequency phonons shorter. The contribution of these low-frequency phonon modes to the thermal conductivity is further limited by the reduction of group velocity and lifetime.

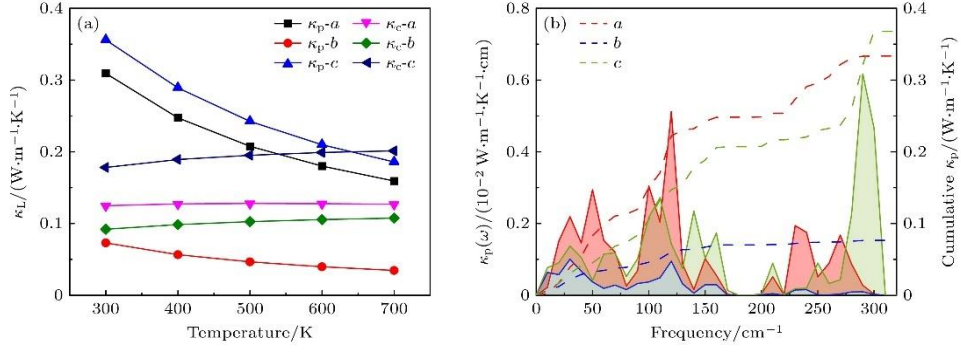

**Figure 4.** (a) Temperature-dependent lattice thermal conductivity of $AgBiSCl_2$; (b) variation of thermal conductivity and cumulative thermal conductivity with frequency at 300 K.

The parameters affecting the thermal transport properties of $AgBiSCl_2$ were studied to explore the reasons for its low thermal conductivity, and the results are shown in the Fig. 5. The Fig. 5(a) is the group velocity, and it can be seen that the group velocity is mainly below 2 km/s. The low group velocity can be attributed to the mostly flat phonon dispersion relation. Then the effect of temperature on the Grüneisen constant ($\gamma$) is studied, such as the Fig. 5(b), the $\gamma$ is as high as 20 in the low frequency region, indicating the strong anharmonicity of the material. The $\gamma$ of the acoustic branch is larger than that of the optical branch, which has a good suppression effect on the phonon heat transfer [43]. The relationship between phonon lifetime and frequency at finite temperature is shown in Fig. 5(c). The phonon lifetime is concentrated in the range of 0.1-10 ps, and decreases with the increase of temperature. A large number of phonons in Fig. 5(c) exist below the Wigner limit, indicating that wave-like phonon tunneling plays an important role in heat transport.

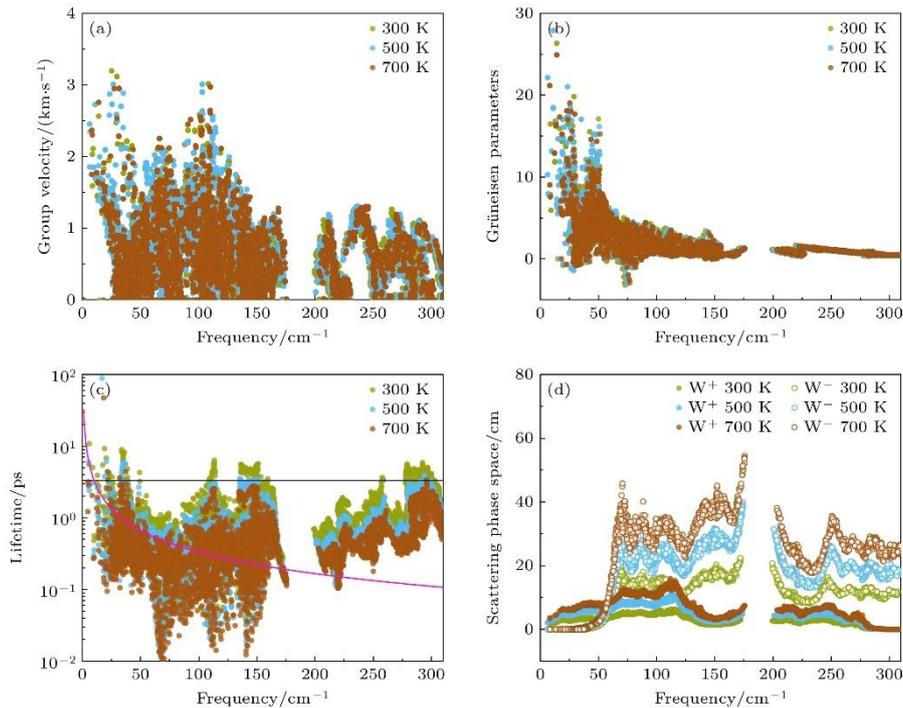

**Figure 5.** (a) Group velocity; (b) Grüneisen parameter; (c) phonon lifetime (black denotes the Wigner limit, purple denotes the Ioffe–Regel limit); (d) three-phonon scattering phase space.

In order to obtain the cause of phonon lifetime reduction due to phonon anharmonic renormalization, the three-phonon scattering phase space is further analyzed. The number of scattering channels (i.e., phase space) available to all phonons in a three-phonon scattering process can be measured by the weighted phase space W. $W^+$ is the absorption process and $W^-$ is the emission process. As shown in the Fig. 5(d), the absorption process dominates in the low frequency region, while the emission process dominates in the high frequency region. Taking into account the phonon anharmonic renormalization, $W^+$ and $W^-$ increase significantly with increasing temperature. Multiphonon interaction induces phonon migration in the low frequency region, which reduces the coupling strength between acoustic and optical modes in the low frequency region, thus reducing the phonon lifetime.

Fig. 6 shows a three-dimensional visualization of the coherent thermal conductivity $\kappa_c$ ($\omega_{qj}$, $\omega_{qj'}$). Quasi-degenerate eigenstates ($\omega_1 \approx \omega_2$) contribute most to $\kappa_c$, because the smaller the energy difference between the eigenstates, the stronger the wave-like tunneling effect. And the contribution of the low frequency region is large, because the phonon band gap in this region is relatively smaller and the mode distribution is dense, so it is easier to stimulate coherent transport. Especially for the phonons involving the rattling vibration of Ag atoms in the low frequency region, the strong anharmonicity leads to enhanced phonon broadening, so the effect of $\kappa_c$ cannot be ignored in the calculation of the thermal conductivity of AgBiSCl$_2$.

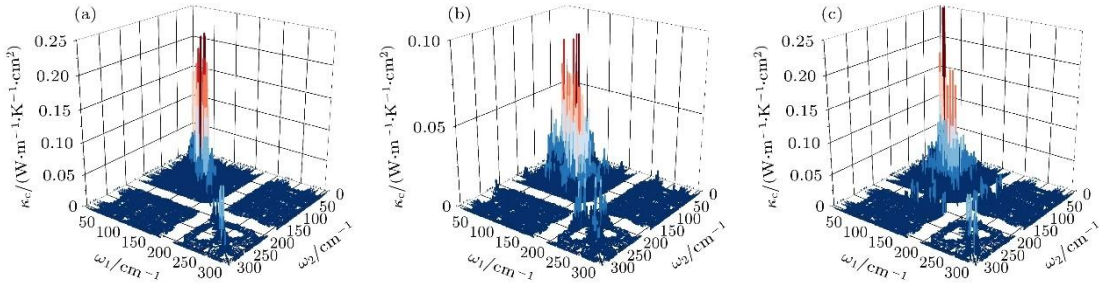

**Figure 6.** Three-dimensional visualization of mode-specific contributions to coherent thermal conductivity along (a)−(c) the *a*, *b*, and *c* directions at 300 K.

The Fig. 7 is the electrical transport parameter of the AgBiSCl$_2$ at temperatures of 300, 500 and 700 K for both n-type and p-type doping in each direction. The Fig. 7(a) and (d) show the variation of Seebeck coefficient with carrier concentration. At doping concentration, the Seebeck coefficient of p-type AgBiSCl$_2$ is larger than that of n-type, which is due to the degeneracy of energy levels near the valence band, resulting in a larger density of States than that near the conduction band. From the Fig. 7(b) and (e) , it can be seen that the conductivity of n-type and p-type AgBiSCl$_2$ increases with the increase of carrier concentration, and the growth rate of n-type is much larger than that of p-type. This indicates that the conductivity of n-type AgBiSCl$_2$ is more sensitive to the increase of carrier concentration.

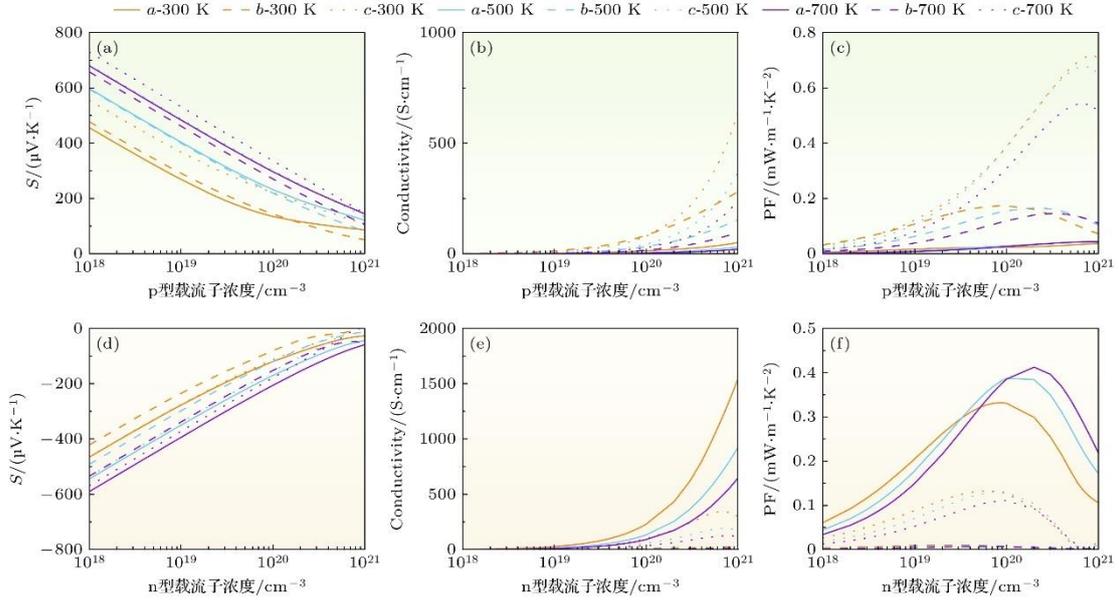

**Figure 7.** Electrical transport parameters of AgBiSCl$_2$ as a function of carrier concentration: (a), (d) Seebeck coefficient; (b), (e) electrical conductivity; (c), (f) power factor.

Fig. 7(c), (f) are curves of power factor (PF) versus carrier concentration at different temperatures. For p-type AgBiSCl$_2$, the PF along the $c$ axis is larger than that along other directions, and the maximum PF can reach 0.72 mW/mK$^2$; for n-type AgBiSCl$_2$, the PF increases with the increase of carrier concentration, and then decreases gradually, and the maximum PF is 0.41 mW/mK$^2$ along the $a$ direction, corresponding to a concentration of 2× 10$^{20}$ cm$^{-1}$.

### 3.4 Electrical transportability

The $ZT$ of AgBiSCl$_2$ is shown in the Fig. 8. The $ZT$ values of n-type and p-type AgBiSCl$_2$ increase first and then decrease with the increase of carrier concentration. The maximum $ZT$ values in different directions are different, and the corresponding carrier concentrations are also different. They have anisotropic thermoelectric properties. For example, Fig. 8(a) and(b), at 700 K, the carrier concentrations are 5×10$^{20}$ and 8×10$^{19}$ cm$^{-3}$, respectively, and the maximum $ZT$ values of p-type ($c$ direction) and n-type ($a$ direction) doping are 0.77 and 0.69. The $ZT$ value of p-type AgBiSCl$_2$ is larger than that of n-type, so AgBiSCl$_2$ can be used as a p-type thermoelectric material. It is worth noting that Chen $et$ $al.$[44] studied p-type LiCu$_3$TiS$_4$ and LiCu$_3$TiSe$_4$, calculated that the $ZT$ reached 0.8 at 700 K, which was close to that in this paper. The thermal conductivity was reduced by applying 3% strain, and finally the $ZT$ value was further increased to 1.6, which pointed out the direction for the subsequent optimization of the thermoelectric properties [44].

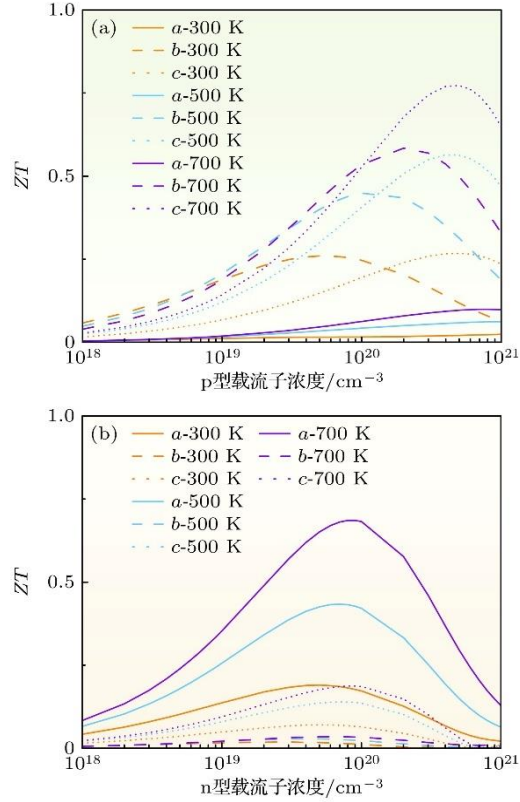

**Figure 8.** *ZT* of AgBiSCl$_2$ at different temperatures: (a) p-type doping; (b) n-type doping.

## 4. Conclusion

In this paper, the photoelectric and thermoelectric properties of mixed anion sulfur halide AgBiSCl$_2$ were systematically studied based on first principles. AgBiSCl$_2$ obtained by HSE06-SOC method is a direct band gap semiconductor with a band gap of 1.72 eV. The AgBiSCl$_2$ has an absorption coefficient of $1\times10^6$ cm$^{-1}$ in the ultraviolet region. The spectral limit maximum efficiency is 28.06% at the thickness of 3 μm, showing a good potential for optoelectronic applications. In AgBiSCl$_2$, the bonding between Ag atom and the surrounding S and Cl atoms is weak, and the local structure is soft, which may lead to strong anharmonicity. It is further found that the Ag-dominated low-frequency local "rattling" vibration leads to the anharmonic scattering effect of phonons, which significantly increases the Gr Grüneisen constant and reduces the phonon lifetime and group velocity, and stimulates the obvious wave-like thermal transport behavior. In the two-channel heat transport model, the $\kappa_p$ decrease and the $\kappa_c$ increase with the increase of temperature. Both $\kappa_c$ and $\kappa_p$ exhibit extremely low lattice thermal conductivities, which are 0.246 W/(m·K) and 0.132 W/(m·K) for $\kappa_p$ and $\kappa_c$, respectively, at 300 K. The extremely low thermal conductivity means that it may have good thermoelectric properties, and further combined with the electrical transport results, the maximum *ZT* of 0.77 (p-type) and 0.69 (n-type) are obtained at 700 K. To sum up, AgBiSCl$_2$ has both excellent photoelectric conversion ability and low thermal conductivity, and is a multifunctional material with potential for both thermoelectric and photoelectric applications, which is worthy of further exploration and engineering optimization. At present, excellent photoelectric properties have been obtained by theoretical calculation, but its thermoelectric properties need to be further optimized, and the figure of merit *ZT* has not yet exceeded 1, which greatly limits its practical application. In the future, the thermoelectric properties of the material will be improved through strain engineering, energy band engineering, doping and other control methods to achieve a higher figure of merit. It will promote new ideas for the combination of photoelectric and thermoelectric applications of the material and promote the development and application of the material in the field of energy.